\documentclass[prb,
twocolumn,
superscriptaddress,showpacs,amsmath,amssymb]{revtex4}
\usepackage{amsfonts}
\usepackage{bm}
\usepackage{verbatim}

\usepackage{graphicx}

\begin{document}

\title{Topology of the $^3$He-A film on corrugated graphene substrate}

\author{G.E. Volovik}

\affiliation{Low Temperature Laboratory, Aalto University, P.O. Box 15100, FI-00076 AALTO, Finland}

\affiliation{ L.~D.~Landau Institute for
Theoretical Physics, 117940 Moscow, Russia}

\affiliation{P.N. Lebedev Physical Institute, RAS, Moscow 119991, Russia}

\date{\today}

\begin{abstract}
{ 
Thin film of superfluid $^3$He on a corrugated graphene substrate represents topological matter with
a smooth disorder. It is possible that the atomically smooth disorder produced by the corrugated graphene does not destroy the superfluidity even in a very thin film, where the system can be considered as quasi two-dimensional topological material.  This will allow us to study the effect of disorder on different classes of the $2+1$ topological materials: the chiral  $^3$He-A with intrinsic quantum Hall effect and the time reversal invariant planar phase with intrinsic spin quantum Hall effect. In the limit of smooth disorder, the system can be considered as a Chern mosaic -- a collection of domains with different values of Chern numbers.
In this limit, the quantization of the Hall conductance is determined by the percolated domain, while the density of the fermionic states is determined by the edge modes on the boundaries of the finite domains. 
This system can be useful for the general consideration of disorder in the topological matter.
}
\end{abstract}

\maketitle



\section{Introduction}

\begin{figure}
\centerline{\includegraphics[width=1.0\linewidth]{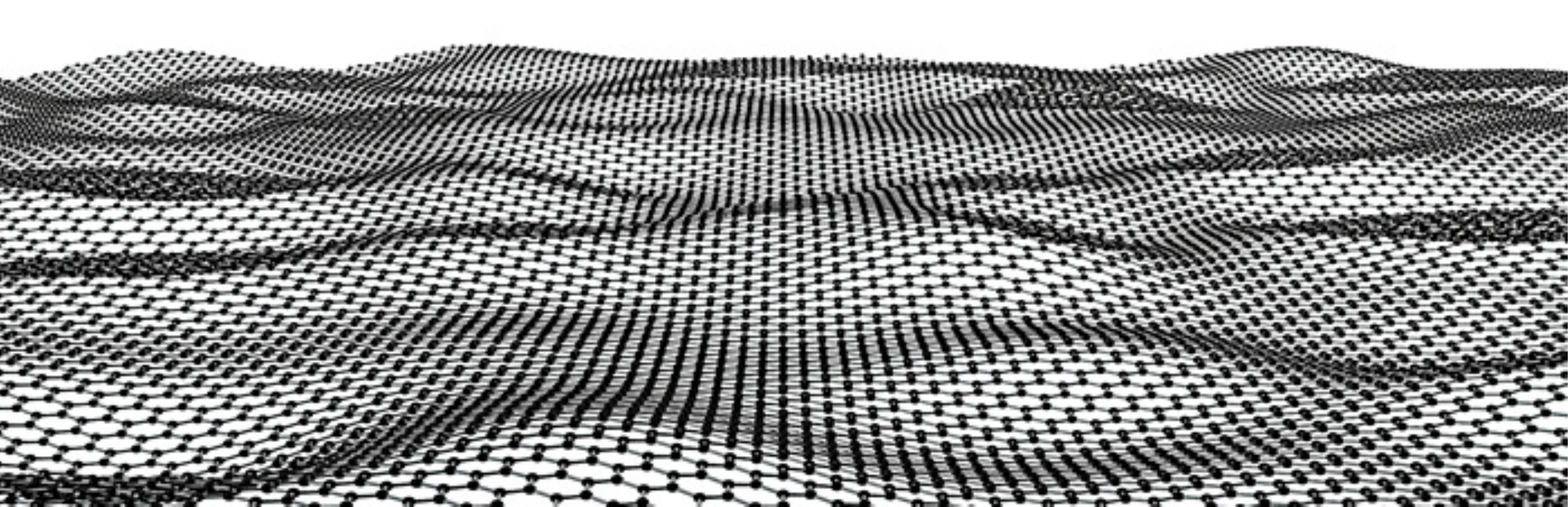}}
  \caption{Corrugated graphene} 
\label{Corrugated}
\end{figure}

Graphene and topological phases of superfluid $^3$He have many
common properties. In particular, in both systems quasiparticles behave as relativistic fermions, being protected by symmetry and topology. These two systems can be combined to produce new phenomena.\cite{KatsnelsonVolovik2014}
Using the thin film of the chiral superfluid $^3$He-A and time reversal invariant planar phase deposited on graphene substrate, one may study the effect of disorder in these $2+1$ topological systems, which experience  the intrinsic quantum Hall and intrinsic spin quantum Hall effects respectively.\cite{Volovik1992a,MakhlinSilaevVolovik2014}
The property of graphene, which is important for this purpose, is the following. On one hand it can be made  atomically smooth, which is the necessary condition for the superfluidity in thin films of $^3$He. On the other hand it is corrugated (see Fig. \ref{Corrugated}), which provides disorder for the $^3$He film.
Let us consider the effect of disorder on $2+1$ chiral topological superfluid $^3$He-A on a corrugated graphene substrate in Fig. \ref{Corrugated}.

On a flat graphene sheet, the film is homogenous, and its state is characterized by the integer valued topological invariant (Chern number) expressed in terms of Green's function ${\cal G}(p_x,p_y,\omega)$:
 \begin{equation}
\tilde N_3 = \frac{1}{4\pi^2}
{\bf tr}\left[\int   d^2p d\omega
~ {\cal G}\partial_{p_x} {\cal G}^{-1}
{\cal G}\partial_{p_y} {\cal G}^{-1} {\cal G}\partial_{\omega}  {\cal
G}^{-1}\right]
\,.
\label{2DN3}
\end{equation}
Here the momentum integral is over the whole momentum space in the $^3$He-A and corresponds to the  homotopy $\pi_3(S^3)$, since the Green's function has the same zero value at infinity. For electron system in solids  the  momentum integration is over the Brillouin zone -- the torus of crystal quasimomentum. The topological charge in Eq.(\ref{2DN3}) determines the quantized Hall conductance in charged $^3$He-A:\cite{Volovik1988,Volovik1992a}
 \begin{equation}
\sigma_{xy}=\frac{1}{2} \,\frac{e^2}{h} \,\tilde N_3 
\,.
\label{conductance}
\end{equation} 
This is twice smaller than for the non-superconducting chiral systems, where $\sigma_{xy}= \tilde N_3 (e^2/h)$.
\cite{So1985,IshikawaMatsuyama1986,IshikawaMatsuyama1987}
The reason for the extra factor $1/2$ in superfluids and superconductors is the Majorana character of the Bogoliubov quasiparticles, with $a^\dagger_{\bf k}=a_{-{\bf k}}$.
 Note that in the relativistic theories the massive $2+1$ Dirac fermions are marginal, because their momentum space is not compact. As a result their topological charge is fractional, 
 $\tilde N_3=1/2\, \rm{sign}\{m\}$, where $m$ is the mass.\cite{Volovik2003} So they also experience the fractional conductance 
 $\sigma_{xy}= \rm{sign}\{m\} \,e^2/2h$ (the compactification by regularization at large momenta makes 
$\tilde N_3$ integer).
The topological charge in Eq.(\ref{2DN3}) also determines the number of the topological edge states.\cite{Volovik1992b}

\begin{figure}
\centerline{\includegraphics[width=1.0\linewidth]{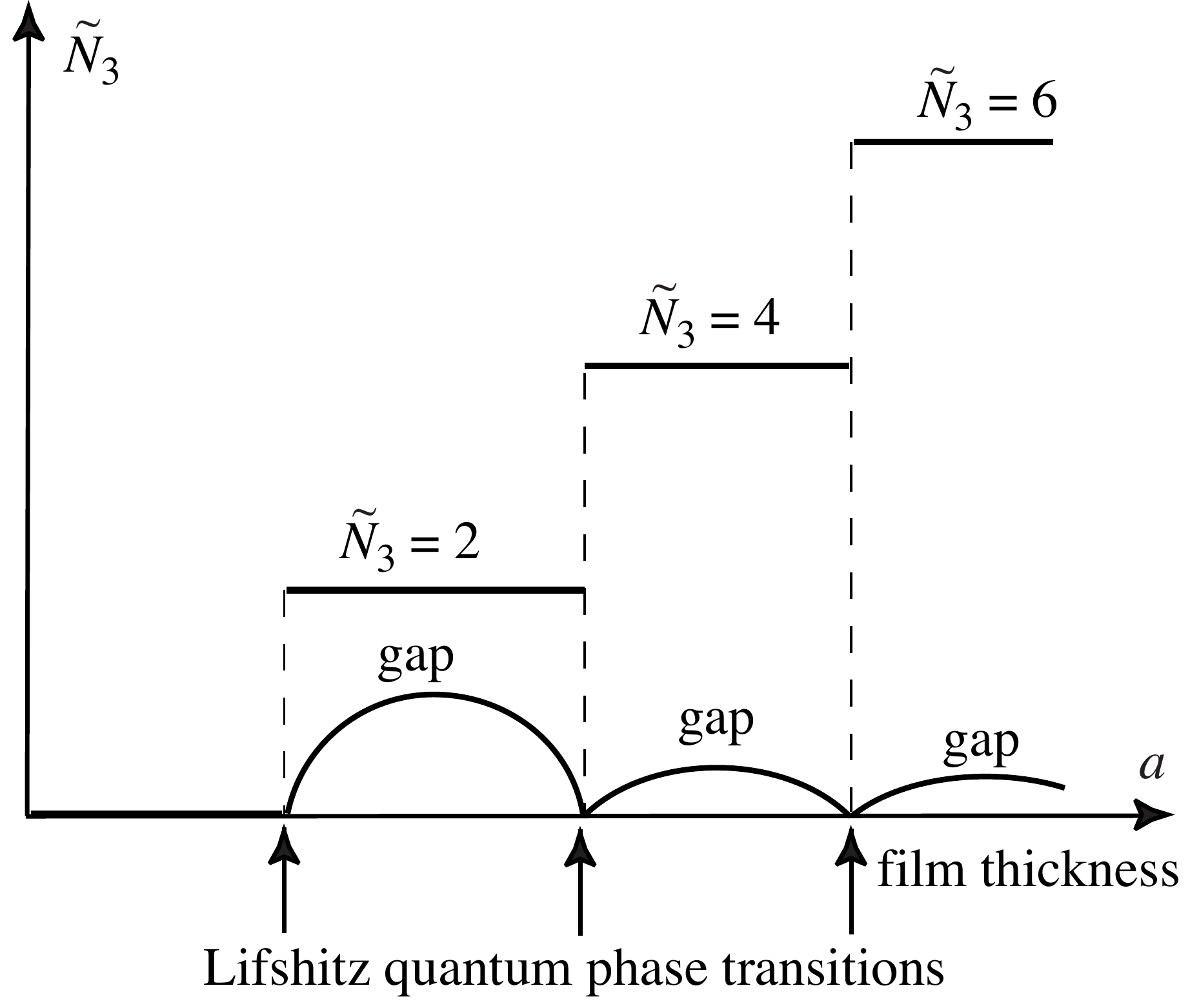}}
\caption{Integer topological invariant $\tilde N_3$ as a function of the
film thickness $a$. In $^3$He-A this Chern number takes even values because of spin degeneracy in the spin triplet superfluids, while in the spin polarized $^3$He-A$_1$ it may take any integer value. Points where $\tilde N_3$ changes abruptly are  topological 
phase transitions -- a kind of quantum Lifshitz transitions. Solid curves show the gap in the  quasiparticle spectrum -- the minimum of the energy. The spectrum becomes gapless at the Lifshitz  transition. The maximun of the gap is proportional  to $\Delta_0/\sqrt{|\tilde N_3|}$, where $\Delta_0$ is the gap amplitude in bulk $^3$He-A. In the limit of large thickness of the film, $|\tilde N_3|\rightarrow \infty$, the gapless bulk state with  the Weyl nodes is formed.}
 \label{N3}
\end{figure}

In the $^3$He film, the topological invariant $\tilde N_3$ depends on the film thickness $a$, see Fig. \ref{N3}. For atomically thin films, $\tilde N_3$ is proportional to the number of atomic layers.
With increasing of the thickness towards the bulk limit the Weyl points in the spectrum are formed. In this $3+1$ limit, at each value of the momentum $k_z$ the system represents the $2+1$ superfluid. In the interval $|k_z|<k_F$ between the Weyl nodes these superfluids are topolically nontrivial  with  the charge $\tilde N_3=2$. As a result the total Chern number
of the thick film is determined by the splitting $2k_F$ of the Weyl nodes: $\tilde N_3 = 2(2k_F a/2\pi)$, see also \cite{Burkov2017}. In this limit, the topological edge states form the flat band with zero energy in the whole interval $-k_F < k_z < k_F$.\cite{KopninSalomaa1991,Volovik2011}

\section{Topological domains}

\begin{figure}
\centerline{\includegraphics[width=1.0\linewidth]{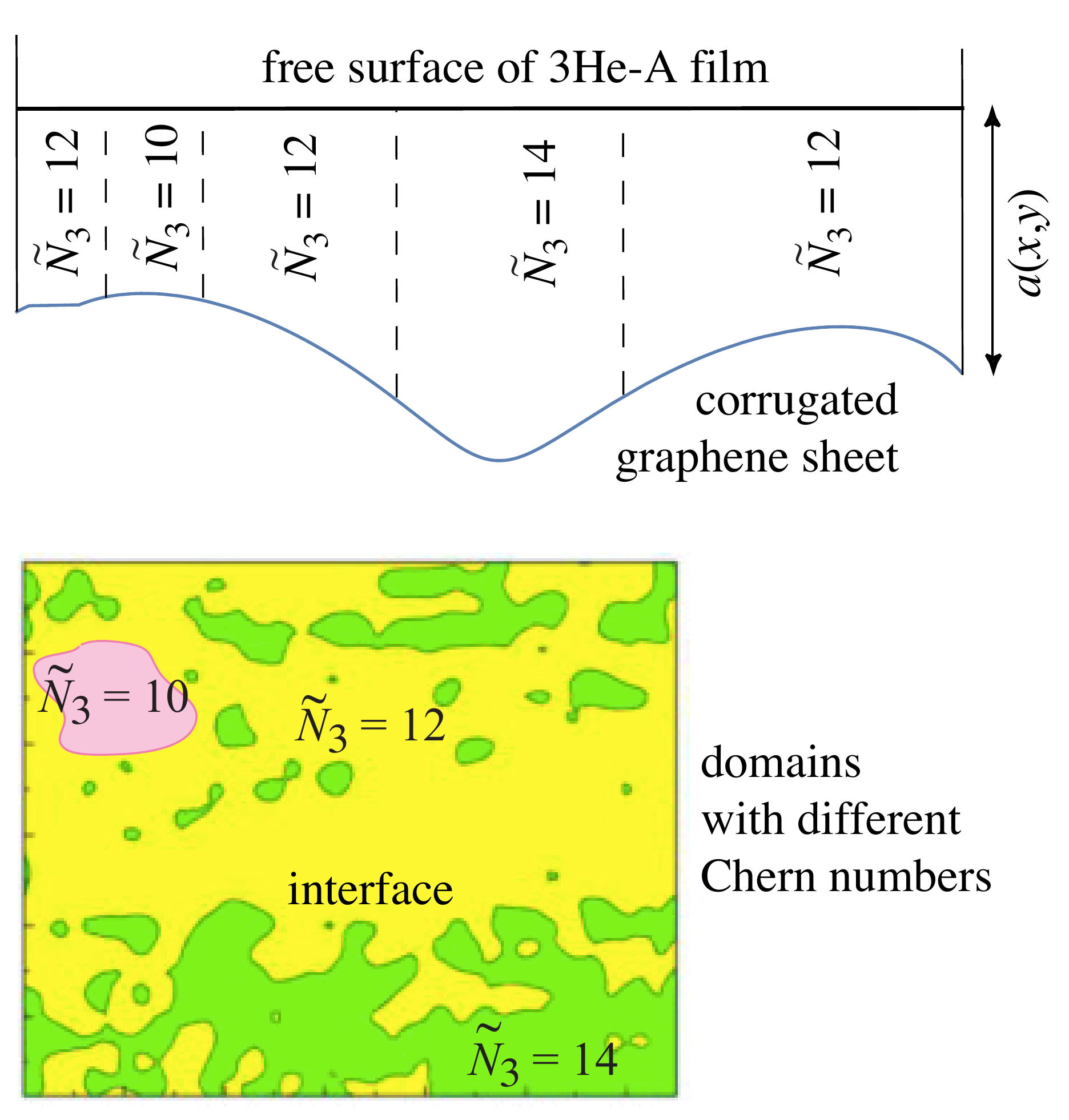}}
  \caption{\label{boundary} 
Illustration of the distribution of the values of topological invariant  in the $^3$He-A film on the corrugated graphene. The local topological  invariant $\tilde N_3(x,y)$ in Eq.(\ref{2DN3}) can be constructed using the local Green's function $G({\bf p},{\bf r},\omega)$. ({\rm top}): The system can be represented as collection of domains with different topological invariants $\tilde N_3$, which is the space analog of the Chern mosaic.\cite{Ojanen2016,Ojanen2017}
The topological invariant of the whole state is determined by the invariant of the infinite (percolated) domain, while the domains with the other values of $\tilde N_3$ are the finite islands.  ({\rm bottom}): This  picture illustrates the configuration with the two infinite domains,  with $\tilde N_3=12$ and  $\tilde N_3=14$.
These two states are separated by the interface of the infinite length, which contains fermion zero mode.
The boundaries of the islands have the finite energy modes with energy inversely proportional to the length $L$ of the boundary of the island.}
\end{figure}

For $^3$He-A on corrugated graphene, the thickness of the film  $a(x,y)$ depends on the coordinates.  Since the corrugation is smooth, one can introduce the local topological charge $\tilde N_3(x,y)$, which can be constructed using the local Green's function $G({\bf p},{\bf r},\omega)$, where ${\bf r}=( {\bf r}_2+{\bf r}_1)/2$ is the center of mass coordinate, and ${\bf p}$ is the Fourier component of the relative coordinate ${\bf r}_2-{\bf r}_1$.   
Fig. \ref{boundary} illustrates the spatial distribution of the local Chern number  $\tilde N_3(x,y)$. The regions with the same value of invariant form either closed islands or the macroscopic percolating cluster. 
The domains with different topological charges are separated by the lines, which contain the edge states.
 The state of the film may contain single percolated domain. The topological charge of such infinite cluster determines the value of the quantum Hall conductance of the whole sample according to Eq.(\ref{conductance}). The current (or the spin current in case of the planar phase) is formed by the delocalized    edge states on the boundaries of the graphene sheet.

The boundaries of the  finite islands also contain the edge states, but these states are localized. Since the length of such boundary is finite, these edge state represent only approximate
zero modes. On the other hand the boundary of the dominating domain has infinite length, and thus contains exact zero modes, whose number $\tilde N_3$ determines the integer valued conductance, see the next section.

The state of the film may contain two percolated domains, see Fig.\,\ref{boundary}. They are separated by the infinite interface with exact zero modes, whose algebraic number equals $\tilde N_3^{(2)}-\tilde N_3^{(1)}$, due to counterpropagating modes.

\section{Edge modes and the density of states}

Both the infinite interface between the percolated domains with different $\tilde N_3$, and the  boundary of the percolated domain have the topologically protected branches of edge states which cross the zero energy level. On the semiclassical level, the infinite boundary contains lines of zeroes of the inverse Green's function  $G^{-1}({\bf p},{\bf r},\omega)$ in the 5-dimensional  $(p_x,p_y,x,y,\omega)$-space. These nodal lines 
are described by the topological charge $N_3$, where the integral is over the $S^3$ spheres around the line. According to the Atiyah-Bott-Shapiro construction,\cite{Horava2005}  the effective Hamiltonian in the vicinity of the line  has the form:
 \begin{equation}
H=\sigma_x c_xp_x + \sigma_y c_yp_y + q\sigma_z x \,,
\label{Hamiltonian}
\end{equation}
where the coordinate $y$ is locally along the line, and the variables $(x,p_x,p_y)$ are perpendicular to the line.
On the quantum level the Hamiltonian has the Majorana fermion zero mode with the linear spectrum 
$E=\tilde c_yp_y$, which crosses zero as a function of $p_y$.\cite{Volovik2003}

The zero energy state exists only on the infinite boundary of the domain or at the infinite  interface between the domains. The finite boundaries of the islands do not have zero modes. However, the energy of the edge states can be arbitrarily small, on the order of $|E| \sim c_y/L$, where $L$ is the length of the boundary of the island. The distribution of the islands and their boundaries with the localized esge states determine the statistical properties of the low-energy fermionic states in the bulk of the system, which can be experimentally studied.

All this can in principle be applied to the electron system with disorder, see e.g. \cite{Chudnovsky1986}. When the disorder is on the quasiclassical level, i.e. when the impurities are smooth and the length scale of the coordinate dependence 
is much larger than the interatomic distance, then one can introduce the local topological  invariant $\tilde N_3({\bf r})$ in Eq.(\ref{2DN3})  using the local Green's function $G({\bf p},{\bf r},\omega)$. 

\section{Discussions}

We considered the $2+1$ topological system with a smooth type of disorder, which can be realized in thin films of superfluid $^3$He on a graphene substrate. The atomically smooth disorder produced by the corrugated graphene should not destroy the superfluidity of the thin film. This will allow us to study the effect of disorder on different classes of the topological materials: the chiral $2+1$ $^3$He-A with intrinsic QHE and the time reversal invariant planar phase with intrinsic spin QHE. In the limit of the smooth disorder the system can be considered as a Chern mosaic --  a collection of domains with different topological Chern numbers.
In this limit, the quantization of the Hall conductance is determined by the macroscopic percolated domain, while the density of the fermionic states in bulk\cite{Movassagh2017} is determined by the edge modes on the boundaries of the finite domains. 
This system can be useful for the general consideration of disorder in the topological matter. \cite{Furusaki2015,Prodan2015,Slager2015,Prodan2016,ShouChengZhang2017,DisorderReview2016}

The bulk 3D topological system with smooth disorder has already been realized in $^3$He-A
 in aerogel. The random anisotropy of the aerogel strands destroys the long range orbital order giving rise to the Larkin-Imry-Ma disordered state.\cite{Volovik2008,Dmitriev2010} This  is a superfluid realization of the skyrmion glass suggested for magnets.\cite{Chudnovsky2017} In the case of the $^3$He-A the skyrmions are 3-dimensional\cite{Skyrme1962} and are described by the Hopf invariant expressed in terms of the superfluid helicity.\cite{Volovik1977,Ruutu1994,Makhlin1995}
In this disordered state the Weyl nodes are smoothly distributed in space and form a unique example of a Weyl glass. The effective (artificial) tetrad field experienced by the Weyl fermions is disordered, $\left<e_a^\mu \right>=0$, forming the analog of the so-called "torsion foam".\cite{Hawking1978,HansonRegge1979}

I thank Yu. Makhlin for the numerous discussions which led to this paper.
I acknowledge support by RSF (\# 16-42-01100).

\end{document}